# Complementary ADF-STEM: a Flexible Approach to Quantitative 4D-STEM

Bryan D. Esser, [a] Joanne Etheridge [a,b]

a. Monash Centre for Electron Microscopy, Monash University, VIC 3800, Australia
b. Department of Materials Science and Engineering, Monash University, VIC 3800, Australia

*Highlights:*
- ADF-STEM images are formed from the complement of low-angle scattered electrons
- Solution to the tradeoff between angular resolution and collection angle in 4D-STEM
- Improved experimental flexibility, including dose-optimization
- Facilitates use of detectors with low dynamic range
- Method for quick, high precision *in situ* beam current measurement

## Abstract
Scanning transmission electron microscopy (STEM) has a broad range of applications in materials characterization, including real-space imaging, spectroscopy, and diffraction, at length scales from the micron to sub-Ångström. The recent development and adoption of high-speed, direct electron STEM detectors has enabled diffraction patterns to be collected at each probe position, generating four-dimensional STEM (4D-STEM) datasets and opening new imaging modalities. However, the limited pixel numbers in these detectors enforce a tradeoff between angular resolution and maximum collection angle. In this paper, we describe a straightforward method for quantifying 4D-STEM data by utilizing the full flux of the electron beam, including electrons scattered beyond the limits of the detector. This enables significantly increased experimental flexibility, including the synthesis of quantitative, high-contrast complementary annular dark field (cADF) STEM images from low-angle diffraction patterns whilst maintaining high angular resolution; as well as the optimization of electron dose and the more effective use of low dynamic range detectors.

## 1. Introduction
The transmission electron microscope has been a critical tool in materials characterization for decades given its ability to image a wide array of materials from the micron to sub-Ångström scale.[1–5] Over the years, advances in both hardware and sample preparation have led to significant improvements in spatial and energy resolution. The introduction of aberration correction [6–13] plus improvements in system stability (effective spatial coherence) has enabled scanning transmission electron microscopy (STEM) to become standard practice for investigating atomic scale phenomena. Quantitative imaging techniques have been developed to maximize the benefit of such atomic resolution images. These most commonly make use of high angle annular dark field (HAADF) STEM, where the contrast mechanism is largely incoherent and strongly dependent on atomic number, often referred to as Z-contrast.[14–22] While there are some hardware and physical limitations to such quantitative imaging, excellent agreement can be made between a well-controlled experiment and accurate simulation (incorporating all relevant scattering processes and measured instrumental parameters).[23–25]

More recently, the development of high-speed direct electron detectors has led to a revolution in imaging for both TEM and STEM.[26–32] For STEM imaging, high dynamic range direct detectors have allowed diffraction patterns to be collected without saturating the transmitted beam, whilst still having enough sensitivity to detect signal at higher scattering angles, which can be several orders of magnitude weaker. Such sensitivity and improvements in readout speed mean that diffraction patterns can now be collected as a function of probe position, resulting in a 4D-STEM dataset (two reciprocal space dimensions: $k_x$ and $k_y$; two real space dimensions: $r_x$ and $r_y$). 4D-STEM has been implemented for some time under a variety of names with previous-generation detectors;[33–36] however, these detectors lacked the combination of speed and/or dynamic range necessary to make the technique widely applicable. It was only recently that detector hardware and computational ability have made such imaging feasible on a routine basis at scan speeds at or above the kHz regime.

The additional information and flexibility afforded by collecting whole diffraction patterns rather than relying on circular or annular integrating detectors has allowed microscopists to employ a variety of new and powerful image formation mechanisms such as ptychography, center of mass (CoM) imaging, Symmetry-STEM, and virtual aperture masking, to name a few (see review by Ophus [37]). Powerful as these new modalities are, conventional annular dark field (ADF) imaging remains an important quantitative imaging mode given the relatively straightforward nature of the contrast mechanism. Ideally, ADF-STEM images would be recorded simultaneously with the whole diffraction pattern. However, this presents a major challenge for several reasons. Firstly, the current-generation of high dynamic range 4D-STEM detectors have relatively few pixels, typically 128x128 or 256x256. This sets up a tradeoff between maximizing the angular resolution (pixel densities) and maximizing the scattering angle (the numerical aperture of the detector) for the recorded diffraction pattern. Secondly, limited mounting options within the microscope can result in some 4D-STEM cameras physically precluding the simultaneous use of conventional integrating detectors that would normally be used to form HAADF-STEM images, requiring separate acquisitions to collect both. Furthermore, since many current-generation 4D-STEM detectors have been installed as microscope add-ons, software integration between conventional and 4D-STEM detectors is not always possible. Finally, a detector with sufficient pixel density to record both a large angular range and high reciprocal space resolution would necessarily demand exceedingly large computational resources for collection and processing.

In this paper we demonstrate a method for generating quantitative higher angle ADF-STEM images from 4D-STEM datasets, without compromising the angular resolution of the diffraction pattern at lower scattering angles. This is achieved by first normalizing 4D-STEM data by beam current, then synthesizing 4D-STEM ADF images from the complement of low-angle scattered electrons. Additionally, we show that these "complementary ADF" (cADF) STEM images can provide significant benefits in absolute quantitative imaging, dose efficiency, and applications using detectors with lower dynamic range.

## 2. Complementary ADF Methodology

In the absence of a specimen, the STEM probe in vacuum should undergo no scattering and thus simply result in an image of the probe-forming aperture on the 4D-STEM camera, assuming a suitable camera length has been chosen (Figure 1a). The total number of counts recorded on the camera, $I_T$, then corresponds to the total beam current where $I_T = 100\%$. By contrast, when the

STEM probe passes through a specimen it may result in scattering beyond the edges of the detector, thus reducing the current measured by the detector ($I_D < I_T$), as seen in Figure 1b. While the angular dependence of the scattering distribution beyond the detector cannot be recovered, the total current scattered beyond the detector can be recovered by simply subtracting the measured current from the total current ($I_D - I_T$). As described below, this simple relationship can be applied to synthesize cADF-STEM images with electrons in the angular range $[\theta: \infty)$ so long as the scattering angle $\theta$ is less than or equal to the numerical aperture of the 4D-STEM detector. In order to do so, the beam current must first be accurately measured.

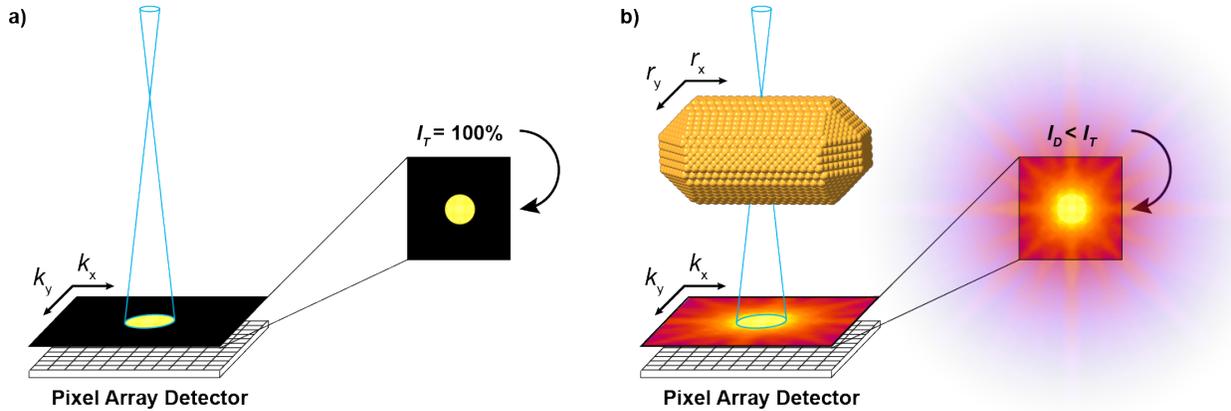

**Figure 1:** A STEM probe in vacuum (a) will result in 100% of the beam current landing on the detector ($I_T$). The presence of a specimen (b) will result in less current on the detector ($I_D < I_T$), where current scattered beyond the detector can be directly recovered as $I_T - I_D$.

2.1. Beam current characterization

Under the desired imaging and detector conditions, an image of the probe-forming aperture is acquired on the pixelated detector with the probe in vacuum (i.e. no specimen). For improved precision, it is recommended that a series of images are captured as the measured beam current may fluctuate due to factors such as source fluctuations or shot noise, especially at lower currents. As an example, a set of $10^4$ images of the aperture in vacuum at 1 ms per image (~10 s total) is summed in the plane of the detector ($k_x$, $k_y$) and the distribution of total counts per frame (shown here in picoamperes) is fitted with a normal distribution (Figure 2). The mean ($\mu$) is taken to be $I_T = 100\%$ beam current at the given imaging conditions, and the relative standard deviation (RSD) is used as a measure of precision, as shown in Equation 1, where $\sigma$ is the standard deviation of the distribution. It is worth noting that the use of RSD as a measure of precision is only valid for data with a true zero reference point, otherwise comparison across different measurements may not be possible.

$$RSD = \frac{\sigma}{\mu} \quad \quad \text{Equation 1}$$

The data presented in Figure 2 have an average beam current of 7.204(4) pA and a standard deviation of 0.0356(5) pA, resulting in an RSD of just 0.49% for the chosen experimental setup (source, gun lens, spot size, condenser aperture, dwell time, etc.). This process was repeated for several beam currents to demonstrate the relationship between RSD and beam current (Figure A1). Additional details for quantifying beam current are also discussed in the Appendix. For an early generation (2007) double spherical-aberration ($C_3$) corrected FEI Titan³ 80-300 with a 5-year-old

SFEG source at 300 kV, the spread in beam current measurements was at the shot noise limit for beam currents less than 10 pA, and slightly higher for larger currents. For reference, beam currents of 1 pA and 10 pA exhibit RSDs of 1.2% and 0.42%, respectively.

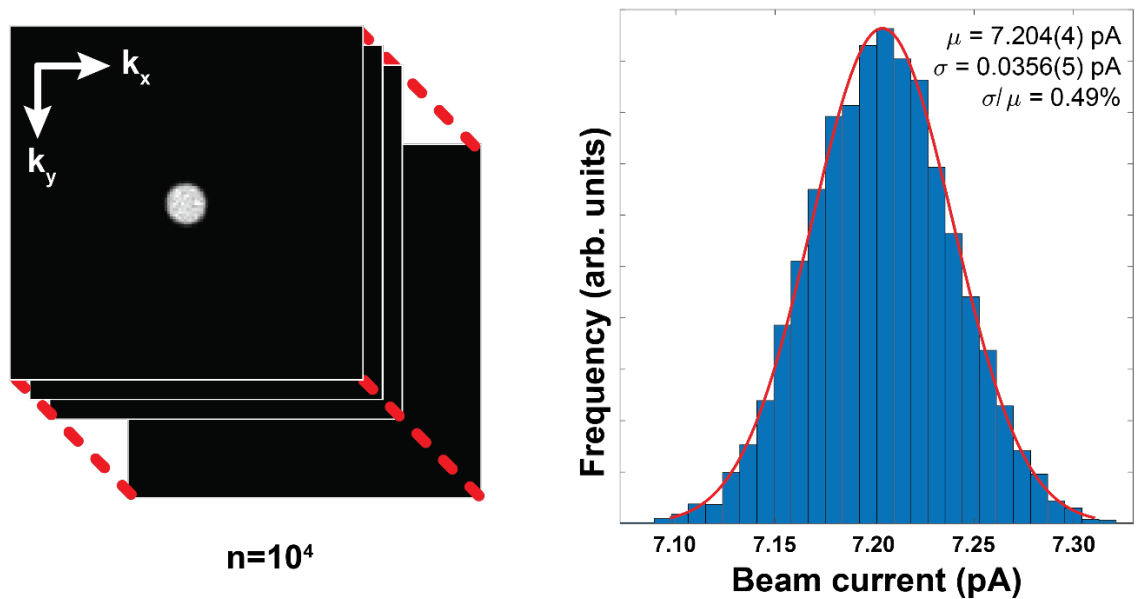

**Figure 2:** Many frames of the condenser aperture in vacuum are summed in ($k_x$, $k_y$) and fit with a normal distribution. Nominal beam current is taken as the mean ($\mu$), with a precision given by the relative standard deviation ($\sigma/\mu$).

In experimental 4D-STEM datasets that include probe positions consisting of vacuum or weakly scattering support material (i.e. ultrathin/lacy carbon, graphene, etc.), the diffraction patterns from such probe positions can be used to calibrate the beam current *in situ*, assuming that a sufficiently large scattering angle is subtended on the detector to ensure all electrons are captured. We confirm that this process works equally well through typical ultrathin and lacey carbon support films as it does through vacuum (see Figure A2), given the very weak scattering caused by such support materials.

2.2. Data normalization
It is convenient to normalize the data by the beam current (or counts) for easy comparison with simulation and across different datasets, though this is not strictly necessary so long as the total counts in the beam current are well-characterized as described above. Figure 3 shows a 4D-STEM dataset collected near the end of a [100]-oriented Au nanorod. The data were acquired with 256x256 probe positions on the FEI Titan[3] 80-300 using an EMPAD direct detector with 128x128 pixels and a convergence semi-angle of 15 mrad at 300 kV. In Figure 3a, a 4D-STEM image formed by summing all counts in reciprocal space, $I(\mathbf{r}) = \sum_{\mathbf{k}} I(\mathbf{k}, \mathbf{r})$, shows the total number of counts collected as a function of probe position, **r**. In this image it can be seen that the total counts collected on an Au column can be significantly less than the total counts collected off the specimen, indicating that a fraction of the beam current (up to 14% in this example) has scattered beyond the reach of the detector.

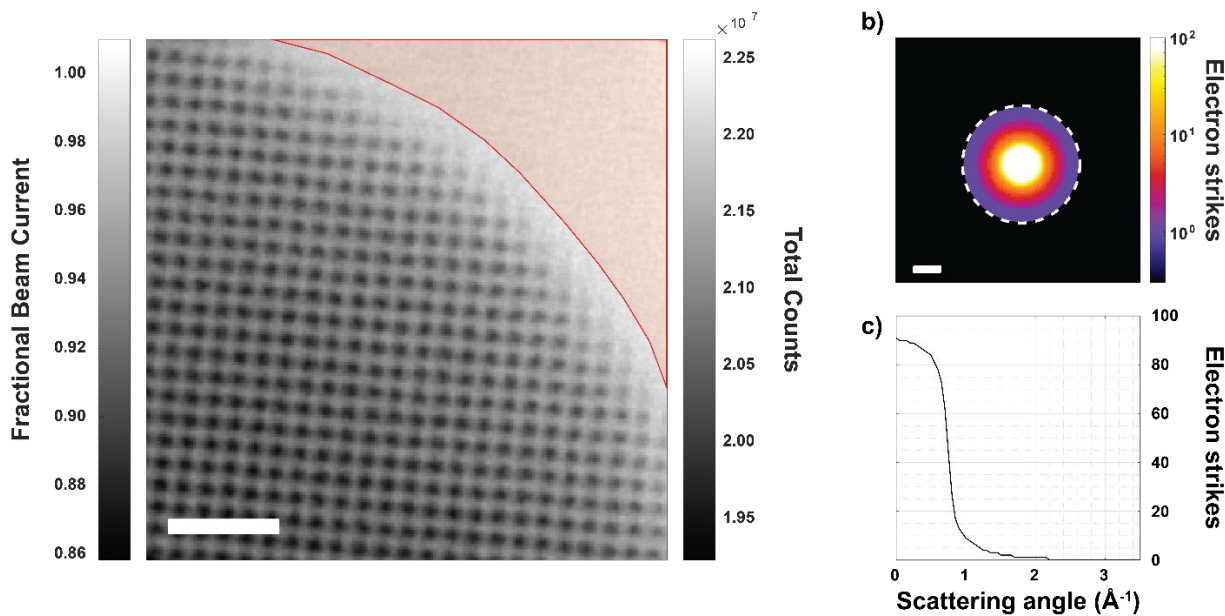

**Figure 3:** (a) 4D-STEM image of [100]-oriented Au nanoparticle showing the total counts collected at each probe position. Diffraction patterns at every probe position are normalized to fractional beam current using the average counts in the area indicated in red, which contains only ultrathin carbon [scale bar: 1 nm]. (b) Position averaged convergent beam electron diffraction pattern from the red area in (a) used for normalization with no electron strikes beyond 2.25 Å$^{-1}$ (44 mrad) [scale bar: 1 Å$^{-1}$], indicated by the white circle and shown in the radial average (c).

The region indicated in red in Figure 3a containing only the ultrathin carbon support film was used for beam current calibration and normalization, where the average counts in the region were taken as 100% beam current. All the scattering in the position averaged convergent beam electron diffraction pattern used for normalization (Figure 3b) occurs within the central half of the detector indicated by the white circle, ensuring that the full beam current was accurately characterized, including the effect of the carbon support. This is also shown in the radial average of Figure 3b shown in Figure 3c, where no electron strikes are recorded beyond $|\mathbf{k}| = 2.25$ Å$^{-1}$ ($\theta = 44$ mrad). Furthermore, the radial distribution of counts beyond the probe forming aperture reflects the true symmetry of the object (unlike the asymmetric response of conventional ADF detectors due to variations in the geometry of photon deflection into the light pipe) and clearly reduces to zero in a quantized fashion.

The RSD of the measured beam current in Figure 3 was 0.52% [6.2440(3) ± 0.0323(4) pA] using $n = 13642$ probe positions in the region indicated in red. The beam-current-normalized intensity scale in Figure 3a shows that after the normalization process some probe positions indicate fractional beam currents greater than unity due to the inherent spread in measured beam current (also demonstrated in Figures 2, A1, and A2).

2.3. Generation of cADF-STEM Images
Using the normalized data in Figure 3, typical STEM images such as bright field (BF: 0-15 mrad), annular bright field (ABF: 10-15 mrad), and annular dark field (ADF: 60-88 mrad) can be directly generated from the experimental data by masking the diffraction pattern with the desired detector geometry, as seen in Figure 4a. Such virtual detectors are limited to the scattering angles that fall within the bounds of the detector, necessitating a tradeoff between angular resolution and maximum scattering angle, as noted above. In this conventional 4D-STEM example, there is sufficient angular range to enable the generation of an ADF image with a relatively high inner collection angle (60 mrad); however, the finite size of the detector results in a smaller than desired outer collection angle (88 mrad) for a true HAADF image and still comes at the expense of fine angular resolution. To avoid this tradeoff, a cADF image, $I_{cADF}(\mathbf{r})$, can be synthesized using all the current scattered beyond a given angle $\theta$ by integrating the 4D-STEM dataset, $I(\mathbf{k},\mathbf{r})$, over the finite angular range $0 \leq \mathbf{k} \leq \theta$ and then subtracting that from unity (representing the normalized beam current), as per Equation 2, yielding a proper HAADF-STEM image.

$$I_{cADF}(\mathbf{r}) = 1 - \int_0^\theta I(\mathbf{k},\mathbf{r})\,d\mathbf{k} = \int_\theta^\infty I(\mathbf{k},\mathbf{r})\,d\mathbf{k} \qquad \text{Equation 2}$$

Figure 4b provides an example of this process where $\theta = 60$ mrad, with the general form of the integration used shown on the righthand side of each image. There are several potential benefits to synthesizing cADF images in this way, depending on the goals of the experiment and the type of detector used. Here we focus on three main benefits including improved contrast and quantification, increased flexibility when using 4D-STEM detectors with lower dynamic range, and improved angular resolution at lower scattering angles. Finally, we show that this normalization process has the added benefit of quick, high-precision *in situ* beam current measurement.

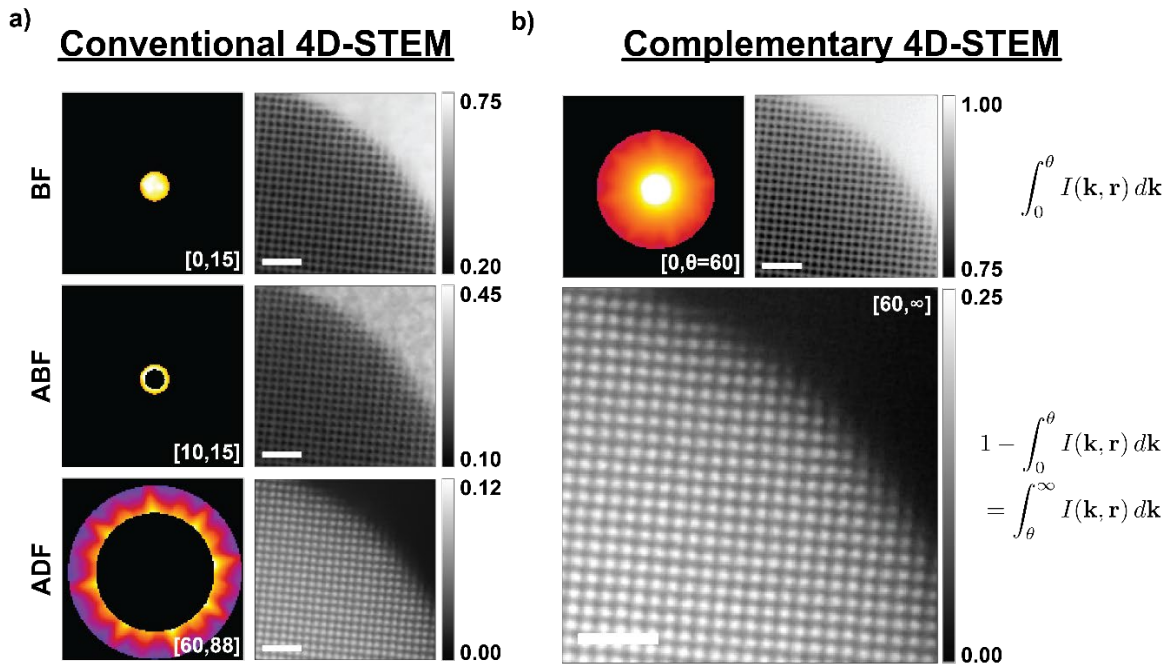

**Figure 4:** Using the data in Figure 3, (a) a selection of standard 4D-STEM images is generated from circular/annular apertures. (b) a complementary ADF-STEM image formed from all current scattered beyond the angle $\theta = 60\ mrad$ can be synthesized by first integrating the diffraction patterns from 0 to $\theta$ mrad to form an incoherent bright field image (top) and then subtracting that from unity (bottom). Scale bars: 1 nm; Intensity scales in fractional beam current; Corresponding annular/circular detector masks indicated in schematics ([min, max] mrad).

# 3. Benefits of Complementary ADF

3.1. Quantitative imaging
Conventional integrating detectors have been extensively characterized and shown to suffer from asymmetric shapes, nonuniform detection efficiency, nonlinear current response functions, and ambiguities in the inner and outer detection angles.[38–42] While these challenges may not significantly alter an ADF-STEM image qualitatively, they can have significant implications for quantitative imaging, even with careful detector characterization. Complementary ADF overcomes these challenges. An accurate and precise normalization is enabled by the current generation of 4D-STEM detectors which have little to no readout noise, providing a true zero reference with single electron sensitivity. Ensuring a proper normalization is critical for quantitative imaging, especially when comparing experimental and simulated data. This normalization then enables an absolute, quantitative measure of *all* the electrons scattered beyond the detector, without limit, avoiding the artefacts of conventional detectors noted above plus the need for calibration of outer aperture angles and uncertainties arising from high-order aberrations or caustics at the periphery of outer-angle apertures. Additionally, as with BF and ABF images synthesized from 4D-STEM data, collecting a diffraction pattern at each probe position removes any axis centering uncertainty present when using circular/annular detectors with fixed mounting positions within the microscope, as such centering can be done to sub-pixel precision via post-processing with 4D-STEM data.

3.2. Dose-efficiency and improved contrast
The cADF image counts *all* of the electrons scattered beyond the chosen inner angle, without limitation. This ensures maximum use of the incident electrons, as well as superior contrast, relative to standard 4D-STEM ADF images (and to a lesser extent conventional ADF images) where the outer angle is constrained by the detector boundary (or, in some cases for conventional ADF images, a physical constraint within the microscope, such as fixed aperture or liner tube). As an example, while the standard 4D-STEM ADF image and cADF image in Figure 4 have the same inner collection angle, the cADF image is not limited in outer angle. The two images are qualitatively quite similar, but the cADF image yields significantly increased fractional beam current as well as improved contrast. This is demonstrated in the line traces taken from both images shown in Figure 5, where the contrast, (the difference between on- and off-column intensity) is nearly 3 times higher in the cADF image than the standard 4D-STEM ADF image. This can be beneficial for quantitative imaging of specimens where the contrast between different columns/atomic species is expected to be quite low, as well as beam-sensitive specimens where higher contrast provides opportunities for lowering dose.

Complementary ADF-STEM is also of benefit in the imaging of thick and/or strongly scattering specimens, where the amount of current scattered out to higher angles can be quite large, as shown in Figure A3. When considerable electron flux is scattered beyond the numerical aperture of the detector, the signal detected in standard 4D-STEM imaging can be very low; whereas, cADF-STEM can provide superior signal to noise to deliver reliable Z-contrast imaging.

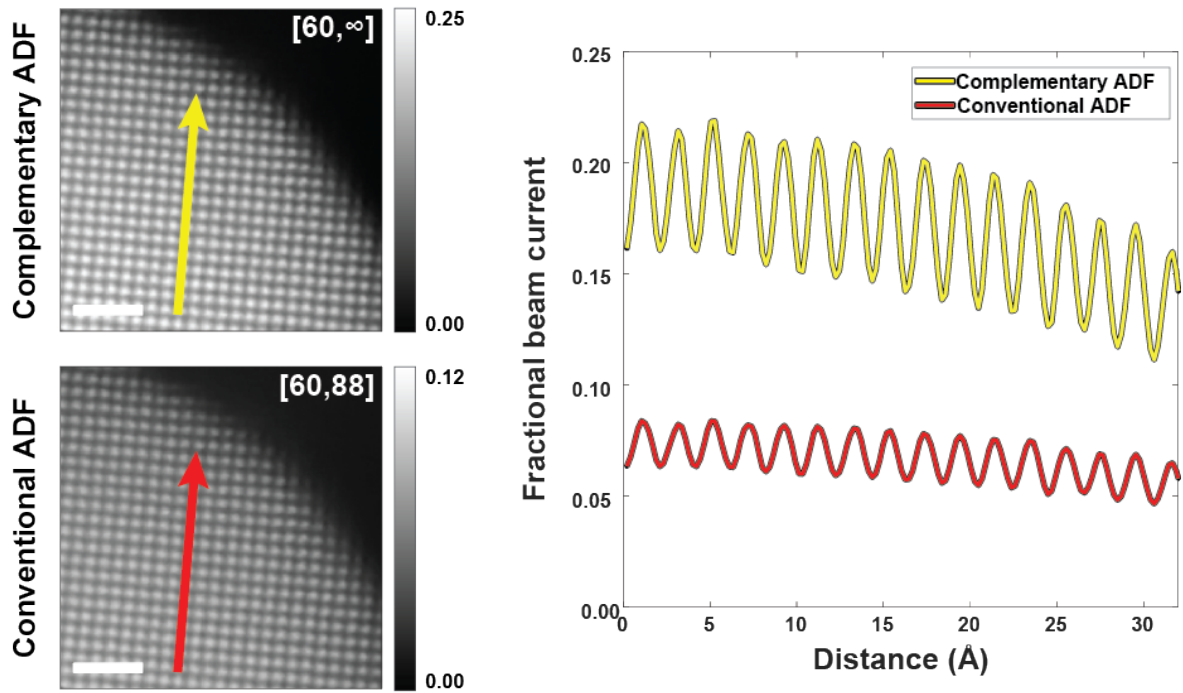

**Figure 5:** Line traces across the cADF (yellow arrow) and conventional ADF (red arrow) images from Figure 4. The cADF image shows increased contrast between atomic columns compared to the conventional 4D-STEM ADF image. Scale bars: 1 nm; scattering angles used for image formation indicated in each image ([min, max] mrad).

3.3. Improved angular resolution and experimental flexibility

There are many imaging modes that use 4D-STEM data that require finer angular resolution at lower scattering angles that can benefit from the simultaneous acquisition of an ADF image. These modes include differential phase contrast, center-of-mass imaging, Symmetry STEM, Lorentz STEM, S-matrix reconstruction, and 4D-STEM ptychography to name a few. [43–47] An ADF image can be used, for example, to determine or validate atomic positions; provide quantitative atomic number and thickness-sensitive contrast; or validate ptychographically reconstructed phase, which often relies on several assumptions about the instrument and specimen.[5,37] Complementary ADF-STEM provides a simple approach for simultaneous acquisition of a quantitative ADF-STEM image without compromising angular resolution or requiring secondary scans using conventional integrating detectors.

3.4. Detector type and experimental flexibility

There are several different direct electron detector types available at present including charge couple devices (CCDs), CMOS monolithic active pixel sensors (MAPS), and hybrid pixel array detectors (PADs). Each detector type trades off different operational capabilities such as maximum frame rate, dynamic range, number of pixels, and/or accelerating voltage compatibility. At 300 kV, the EMPAD used to collect these data can accommodate ~$10^4$ electrons/pixel before saturating at a maximum frame rate of 1.1 kHz.[28] Detectors with different pixel designs can saturate with fewer than 20 electrons at the same frame rate but offer more pixels and/or higher frame rates.[27] Therefore, it is important to ensure that no pixels in the detector are being saturated and any voltage thresholding is properly set, as errors in either/both will skew the beam current measurement.

Given the dynamic range of the EMPAD used here, this would imply very large beam currents or very low camera lengths, such that much of the beam current is focused on just a few pixels. For most 4D-STEM applications, specimen damage precludes the use of such large beam currents, and very low camera lengths would result in poor angular resolution, which is generally counter-productive. For detectors with lower saturation limits, the ability to synthesize cADF images in this way can provide greater flexibility in optimizing experimental setup. For example, longer camera lengths can be used to disperse the beam current across more pixels to avoid saturation. While this results in a lower maximum scattering angle on the detector, which would be problematic for many experiments, the flux scattered to higher angles can still be recovered through the process described above to synthesize cADF images, maintaining increased angular resolution at lower scattering angles for alternative processing methodologies.

3.5. Quick and precise beam current measurement and normalization

The data presented here were collected using an EMPAD direct electron detector, where a linear relationship between counts and electrons can be established. Tate *et al.* report a rough parameterization of 1.97 ADU/keV (ADU: analog-to-digital units).[28] Experimentally, we found the conversion to be 575 ADU at 300 kV, which is close to their parameterized estimate of 591 ADU. While the single electron sensitivity of the EMPAD detector with a 1 ms dwell time yields a theoretical beam current sensitivity of 0.16 fA, in principle this is limited by the shot noise limit, as shown in Figure A1. Still, this makes the process described above, specifically that in Figure 2, a quick and very precise method for characterizing beam current in the electron microscope, that is competitive with other methods.[48]

When such a conversion between counts and electrons is possible and known, the beam current and dose rate are calculated during the data normalization process described above, which are important for experimental repeatability and dose measurement (necessary for beam sensitive materials). While the normalization process does not depend on the data being in either counts or electrons, care should be taken to ensure that the relationship between the two is linear for the given detector.

## 4. Conclusions

We have presented a method for synthesizing complementary ADF images that makes use of electrons that have scattered beyond the reach of the 4D-STEM detector. The technique relies on accurate characterization of the probe current (or counts), which can be done *ex situ* under the same imaging conditions, or *in situ* if a suitable low-scattering region is captured in the 4D-STEM dataset. Once the probe current is known, the dataset may be normalized allowing for the synthesis of cADF images. This is achieved by simply integrating the scattering up to an angle less than the numerical aperture and then subtracting that from unity, the normalized probe current. Such a method may seem rather trivial at first pass; however, it enables increased experimental optimization of angular resolution, current per pixel per frame, and the maximum scattering angle collected. Additionally, it is shown that by counting every electron scattered beyond the chosen inner angle, cADF images maximize dose-efficiency and image contrast. The methods detailed can benefit a range of quantitative STEM techniques including atom counting, center of mass imaging, and ptychography.

## 5. Acknowledgements

The authors thank Dr Weilun Li and Dr Timothy Petersen for their insightful support and helpful discussions and Dr Alison Funston and Dr Anchal Yadav for provision of the Au specimen. This work was supported by Australian Research Council (ARC) grant DP160104679. The authors acknowledge the use of the instruments and scientific and technical assistance at the Monash Centre for Electron Microscopy, a Node of Microscopy Australia, as well as equipment funded by ARC grant LE0454166. We acknowledge the people of the Kulin Nations, on whose land this research was conducted. We pay respect to their Elders, past and present.

## 6. Appendix

Beam current data presented in the manuscript were calculated as described below. Individual frames containing the unscattered beam, such as those in the left-hand side of Figure 2, were summed over ($k_x$, $k_y$). Detector counts on the EMPAD were converted to electron strikes following the protocol described in Tate *et al.*,[28], specifically Figure 5, where the detector is exposed to a low dose of electrons such that any given exposure has sparse, well-separated counts that can be attributed to single or few-electron strikes. When the data are represented as a histogram, discrete quantized peaks representing {0, 1, 2, …, n} strikes are visible. The distance between the peaks was measured to be 575 ADU (analogue-to-digital units) at an accelerating voltage of 300 kV. Thus, all data in counts can be divided by 575 to convert to electron strikes. Finally, the dwell time of 1 ms was used to convert to current, which is a much more common and easily understood representation than detector counts.

The distribution of measured beam currents was fitted using a normal distribution. The reported precision of the standard deviation, $\sigma$, is based on the 95% confidence interval from the fit. The reported precision of the mean, $\mu$, is given as the standard error of the mean, SE, shown in Equation A1, where $n$ is the number of frames used to characterize the beam current.

$$\text{SE} = \frac{\sigma}{\sqrt{n}} \quad\quad \text{Equation A1}$$

In Figure A1, the relative standard deviation in measured beam current is presented for the double aberration corrected FEI Titan$^3$ 80-300 with a 5 year old SFEG source at 300 kV. The data can be fitted by Equation A2 with an R-square value of 0.9988, where RSD is the relative standard deviation in percent and $I$ is the beam current in pA.

$$\text{RSD} = \frac{\sigma}{\mu} \cdot 100\% = 1.297 I^{-0.4932} \quad\quad \text{Equation A2}$$

Given this fit, a beam current of 0.16 fA, representing just 1 electron per 1 ms frame exposure, estimates an uncertainty of 97%. This represents the Poisson noise limit when there is only $N = 1$ event, as the standard deviation, $\sigma$, would also equal 1 ($\sigma = \sqrt{N}$). The deviation from the shot noise limit at beam currents >10 pA is not addressed here as it is still within the hardware specification for the given instrument.

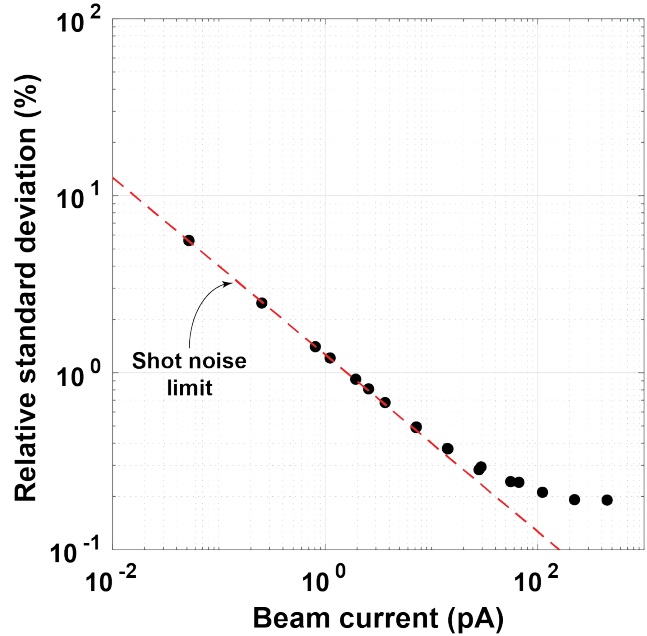

**Figure A1:** Relative standard deviation (in percent) is inversely proportional to the square root of the beam current, following the shot noise limit (red line) for currents <10 pA, slightly increasing for higher currents.

Figure A2 confirms that the normalization process is equally valid *in situ* as it is *ex situ*. The data are collected through lacey carbon support film (a), ultra-thin carbon support film (b), and vacuum (c). While vacuum would naturally be the preferred choice, the thickest of the three, lacey carbon, still does not scatter electrons much beyond the probe-forming aperture. As such, all three are suitable for use in characterization of the total beam current and subsequent data normalization.

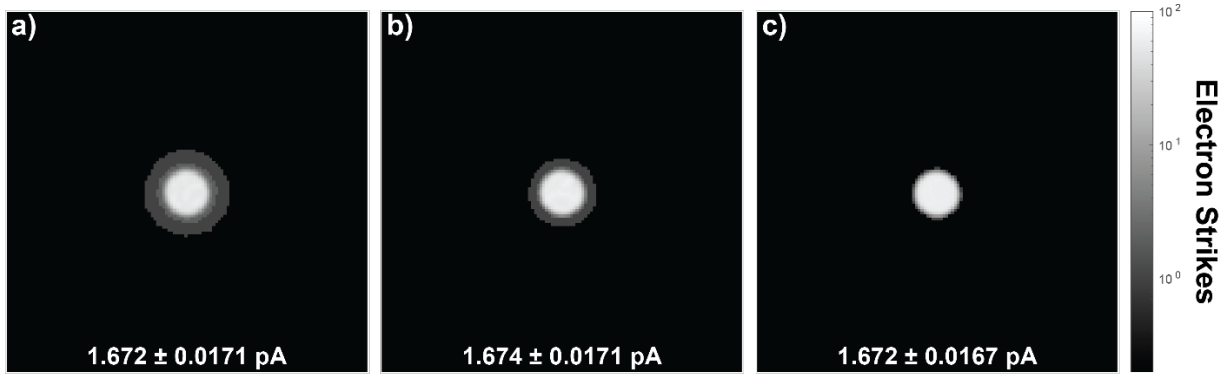

**Figure A2:** Experimental images of a 12 mrad condenser aperture through (a) lacey carbon, (b) ultra-thin carbon, (c) vacuum. Intensity is in electron strikes on a logarithmic scale, where black regions beyond approximately 20 mrad contain zero electron strikes, thus capturing the full beam current for normalization.

Figure A3 demonstrates that the specimen scattering strength and thickness, as well as probe forming aperture size, all affect the amount of beam current scattered through to higher angles, as expected. Simulations using the muSTEM code[49] show that bulk Au scatters a considerably larger fraction of electrons to angles greater than 60 mrad than $SrTiO_3$ does under the same conditions. As such, the ability to synthesize cADF images from 4D-STEM datasets becomes more useful when larger fractions of the beam current are scattered beyond the edges of the detector. Knowing how much of the beam current is expected to be scattered to higher angles can help to better guide experimental setup, making more efficient use of the relatively large doses incurred during typical 4D-STEM experiments, given the slower acquisition rates when compared to conventional integrating detectors.

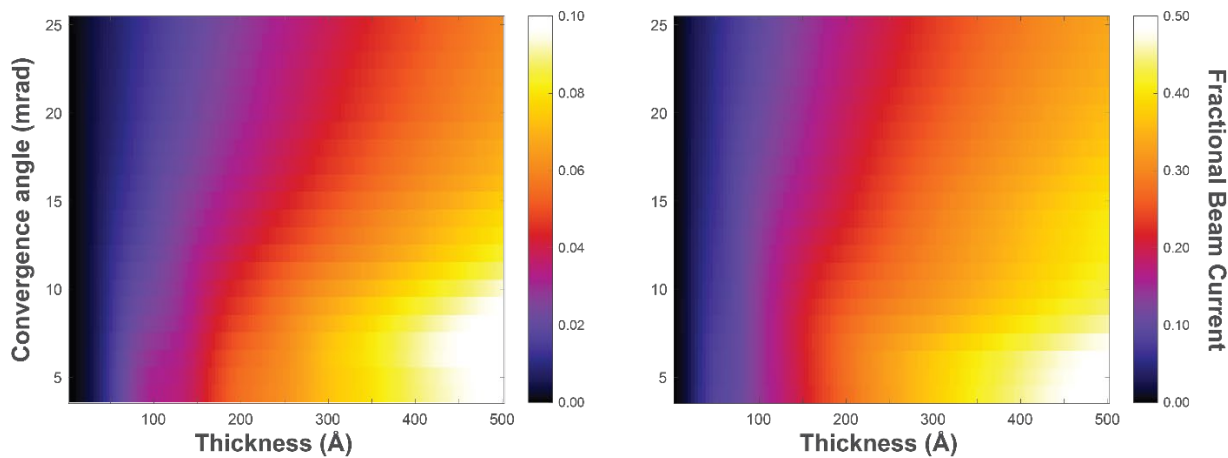

**Figure A3:** Fractional beam current scattered beyond 60 mrad for (left) [100]-oriented SrTiO$_3$ compared to (right) [100]-oriented Au as a function of convergence angle and specimen thickness (all data is simulated). Au scatters significantly more electrons to high angles than SrTiO$_3$, demonstrating the increased benefit of cADF imaging in high-scattering materials.

4D-STEM data presented in the manuscript were collected with a dwell time of 1 ms, which can result in distortions due to nonlinear drift. To improve the quality of the data, two 4D-STEM datasets were collected at the same region with orthogonal scan directions (90° rotation) and then registered via the process described by Ophus *et al.*[50]